\documentstyle[11pt,cs11,html,epsf]{article}

\begin{document}

\title{Chromospheric activity of the double-lined spectroscopic
binary BF Lyn}

\author{D. Montes\altaffilmark{1}, 
R. Lachaume\altaffilmark{2}, 
M.J. Fern\'andez-Figueroa}
\affil{Departamento de Astrof\'{\i}sica, 
Univ. Complutense de Madrid, Spain}



\altaffiltext{1}{Guest observer at McDonald Observatory}
\altaffiltext{2}{Ecole Normale Sup\'erieure, Paris, France}


\setcounter{footnote}{3}


\begin{abstract}

 We present simultaneous spectroscopic observations 
taken  during four observing runs (1996 to 1999) of
H$\alpha$, H$\beta$, H$\epsilon$,  Ca~{\sc ii} H \& K, and Ca~{\sc ii} IRT
lines of the chromospherically active binary BF Lyn.
Both components have strong emission in the H$\epsilon$, Ca~{\sc ii} H \& K and
Ca~{\sc ii} IRT lines and a strong filling-in of the H$\alpha$ and
H$\beta$ lines have
been observed after the application of the spectral subtraction
technique. We have found that the hot component (K2 V)
is always the most active of the system.
The different activity indicators of the hot and cool components
show anticorrelated variations with the orbital phase.

\end{abstract}




%
%
%
\index{*BF Lyn|HD 80715}

\section{Introduction}

  BF Lyn (HD 80715) is a double-lined spectroscopic binary with
spectral types  K2V/[dK] and both components have variable H$\alpha$ emission
and strong Ca~{\sc ii} infrared triplet emission noted by Barden and Nations
(1985). Strassmeier et al. (1989) observed strong 
Ca~{\sc ii} H \& K and H$\epsilon$
emissions from both components. The orbital period is  3.80406 days
(Barden and Nations, 1985) and Strassmeier et al. (1989) from photometric
observations found that BF Lyn  is a synchronized binary with a circular
revolution for a long time.
 Montes et al. (1995) also found strong emission in the Ca~{\sc ii} H \& K
lines from both components with very similar intensities 
and the H$\epsilon$ line
in emission too.

 In this paper we present simultaneous spectroscopic observations of
H$\alpha$, H$\beta$, H$\epsilon$,  Ca~{\sc ii} H \& K, and Ca~{\sc ii} IRT
lines of this chromospherically
active binary. 

\section{Observations}

 Spectroscopic observations in several optical chromospheric activity
indicators of BF Lyn and some inactive stars of similar spectral type and
luminosity class have been obtained during four observing runs.
\newline
1) Two runs were carried out with the 2.56~m Nordic Optical Telescope
(NOT) at the Observatorio del Roque de Los Muchachos (La Palma, Spain) in
March 1996 and April 1998, using the SOFIN echelle spectrograph covering
from 3632~\AA$\ $ to 10800~\AA$\ $
 (resolution from $\Delta$$\lambda$ 0.15 to 0.60 \AA), with a 
1152$\times$770 
 pixels EEV P88200 CCD as detector.
\newline
2) One observing run was obtained using the 2.1~m telescope at McDonald
Observatory (USA) in January 1998 using the Sandiford Cassegrain Echelle
Spectrograph covering from 6382~\AA$\ $  to 8700~\AA$\ $  
(resolution from $\Delta$$\lambda$ 0.13 to
0.20 \AA), and with a 1200$\times$400 pixels Reticon CCD as detector.
\newline
3) The last run was carried out with the 2.5~m INT at the Observatorio
del Roque de Los Muchachos (La Palma, Spain) in January 1999 using the
Multi-Site Continuous Spectroscopy (MUSICOS), covering from 3950~\AA$\ $  to
9890~\AA$\ $  (resolution from $\Delta$$\lambda$ 0.15 to 0.40 \AA), 
with a 2148$\times$2148 pixels
SITe1 CCD as detector.
\begin{figure}
\plotone{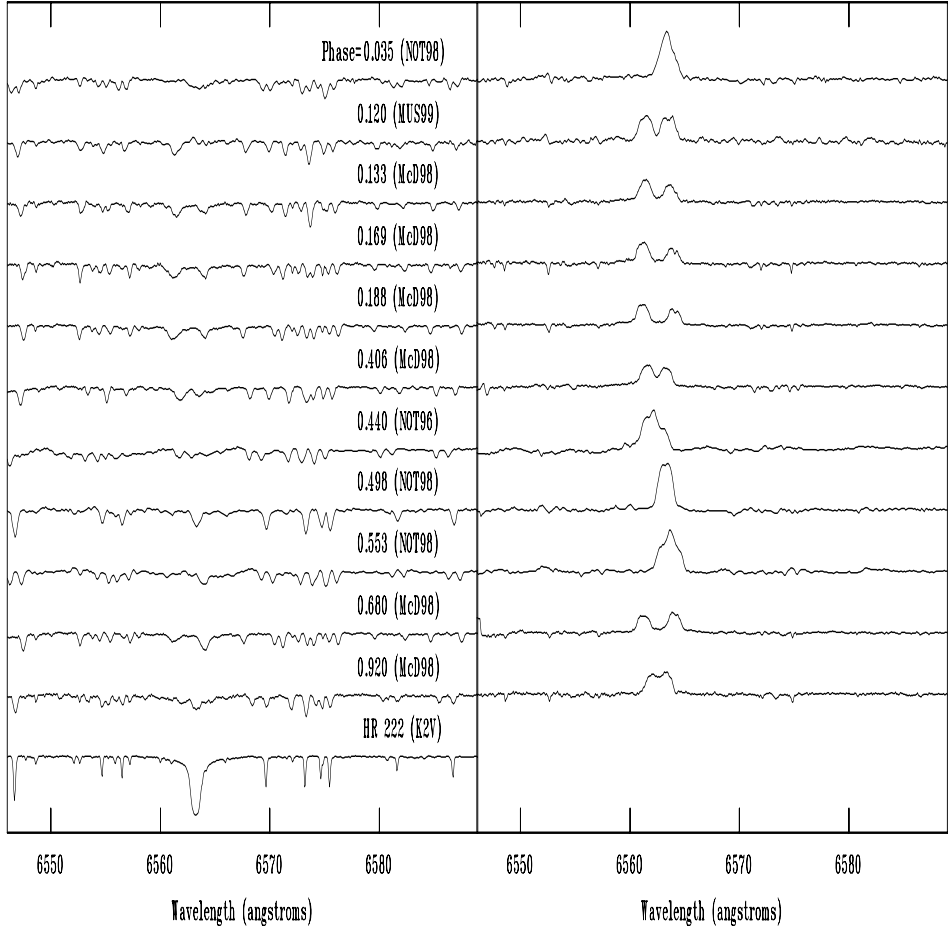}
\caption{The left panel shows the observed spectra of BF Lyn and the
HR 222 (K2V) reference star, in the H$\alpha$ line region.
The right panel shows the subtracted spectra of BF Lyn.} \label{fig-1}
\end{figure}

        In the four runs we have obtained 11 spectra of BF Lyn in different
orbital phases.
Stellar parameters of BF Lyn have been adopted from
Strassmeier et al. (1993), except for T$_{\rm conj}$ taken 
from Barden \& Nations (1985).
        The spectra have been extracted using the standard reduction
procedure in the IRAF package (bias subtraction, flat-field division and
optimal extraction of the spectra). The wavelength calibration was
obtained by taking spectra for a Th-Ar lamp. Finally the spectra have
been normalized by a polynomial fit to the observed continuum.
        The chromospheric contribution in the activity indicators has been
determined using the spectral subtraction technique.

\index{Observatories!Roque de Los Muchachos}
\index{Observatories!McDonald}

%

\begin{figure}
\plotone{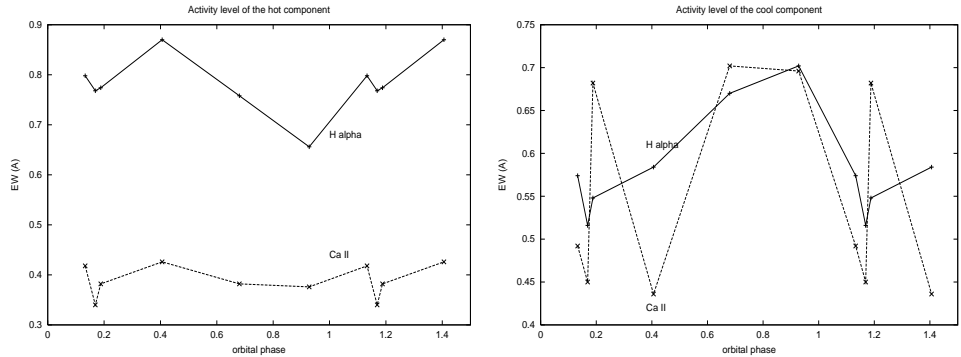}
\caption{H$\alpha$ and Ca~{\sc II} IRT $\lambda$8542 EW of the hot (left panel)
and cool (right panel) components for the McD98 run versus the orbital phase.} \label{fig-2}
\end{figure}

\section{The H$\alpha$ line}

 We have taken several spectra of BF Lyn in the H$\alpha$ line region  in
four different epochs and at different orbital phases. In all the
spectra we can see the H$\alpha$ line in absorption from both components. The
spectral subtraction reveals that both stars have an excess H$\alpha$ emission.
The line profiles are displayed in Fig.~\ref{fig-1}, 
for each observation we plot
the observed spectrum (left panel), and the subtracted one (right panel).
The excess H$\alpha$ emission equivalent width (EW) is measured in the
subtracted spectrum and corrected for the contribution of the components
to the total continuum, in the case of BF Lyn we assume the same
contribution for both stars. 
At some orbital phases, near to the conjunction, is not
possible to separate the contribution of both components.
The excess H$\alpha$ emission of BF Lyn shows
variations with the orbital phase for both components, the hot star is
the most active in H$\alpha$. In Fig~\ref{fig-2} we have plotted for the McD~98
observing run the H$\alpha$ EW versus the orbital phase for the hot and cool
components, respectively. The highest H$\alpha$ EW for the hot component has
been reached at about 0.4 orbital phase and the lowest value is placed at
about 0.9 orbital phase, whereas the cool component shows the highest H$\alpha$
EW at near 0.9 orbital phase and the lowest value between 0.2 and 0.4
orbital phases. The variations of H$\alpha$ EW for both components are
anticorrelated and the  maximum of active areas are found on the faced
hemispheres. The same  behaviour is also found in Ca~{\sc ii} IRT.
  The excess H$\alpha$ EW emission also shows seasonal variations, for
instance, the values of  MUSICOS 99 observing run are very different,
specially for the cool component, from  McD 98 values at  very near
orbital phase.

\begin{figure}
\plotone{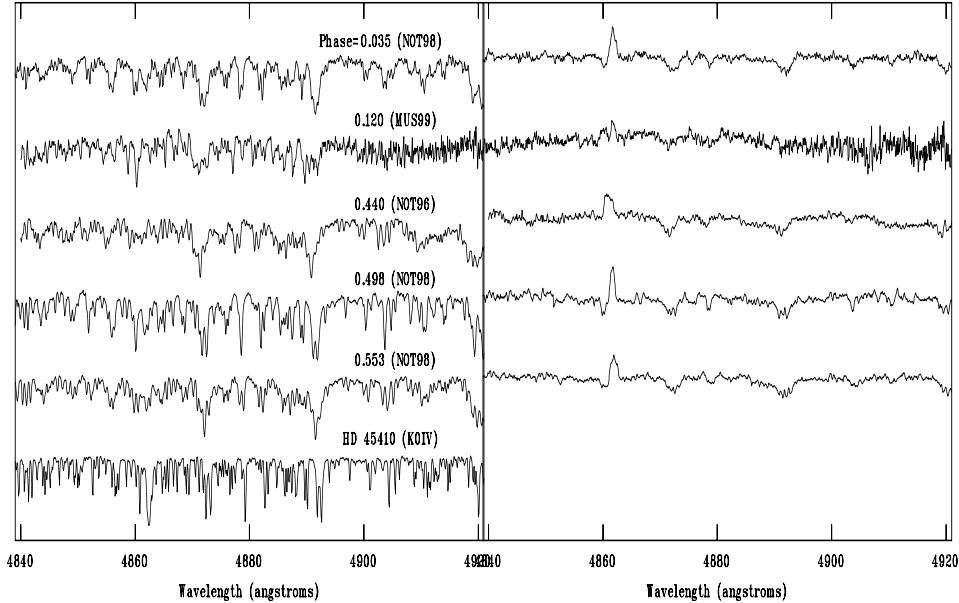}
\caption{As in Fig. 1 for the H$\beta$ line region.}
 \label{fig-3}
\end{figure}

\section{The H$\beta$ line}

   Five spectra in the H$\beta$ region are available in three different
epochs and at different orbital phases. In all the spectra the H$\beta$ line
appears in  absorption from both components, the application of the
spectral subtraction technique reveals a clear excess H$\beta$ emission from
both stars. The line profiles are displayed in Fig.~\ref{fig-3}.
We have determined the excess H$\beta$ emission EW in the subtracted spectra,
the ratio of excess emission EW,  
and the  $ \frac{ E_{H\alpha} } { E_{H\beta} } $ relation:

\begin{equation}
\frac{ E_{H\alpha} }{ E_{H\beta}}=  
\frac{ EW_{sub} (H\alpha) }{ EW_{sub} (H\beta) } 
\cdot 0.2444 \cdot 2.512^{(B-R)}
\end{equation}

given by Hall \& Ramsey (1992) as a diagnostic for discriminating between
the presence of plages and prominences in the stellar surface. The low
ratio that we have found in BF Lyn do not allow us  to discriminate
between both structures.

\section{The Ca~{\sc ii} IRT lines}

\begin{figure}
\plotone{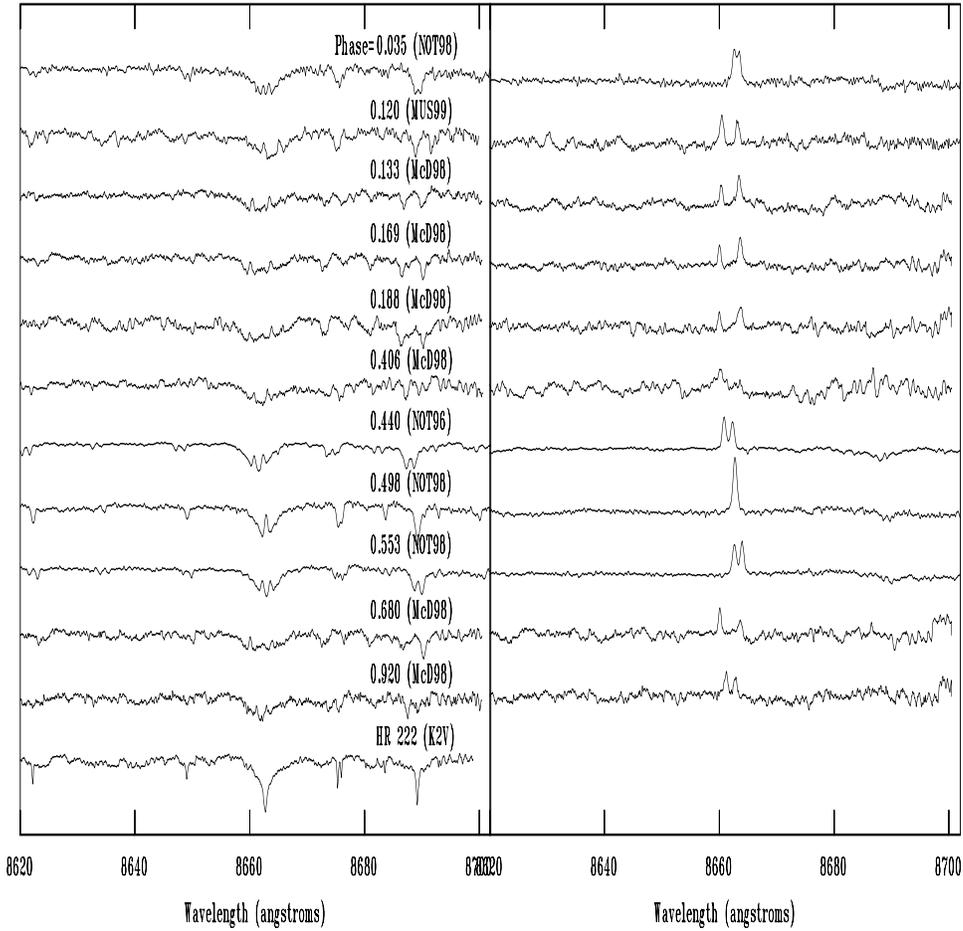}
\caption{As in Fig. 1 for the Ca~{\sc ii} IRT $\lambda$8662 line region.}
 \label{fig-4}
\end{figure}

 The Ca~{\sc ii} infrared triplet (IRT) $\lambda$8498, $\lambda$8542, 
and $\lambda$8662 lines are
other important chromospheric activity indicators. We have taken several
spectra of BF Lyn in the Ca~{\sc ii} IRT lines region in three different
epochs and at different orbital phases, the three Ca~{\sc ii} IRT lines are
only included in MUSICOS 99 observing run. In all the spectra we can see
the Ca~{\sc ii} IRT lines in emission from both components (Fig.~\ref{fig-4}).
As in the case of H$\alpha$ line the 
Ca~{\sc ii} IRT emission shows variations with the orbital phase for
 both components.
In Fig.~\ref{fig-2}  we have plotted, for the McD 98  observing run, 
the Ca~{\sc ii} $\lambda$8542 
EW versus the orbital phase for the hot and cool component,
respectively. The variations of Ca~{\sc ii} emission EW for both components are
anticorrelated and they show the same behaviour found in the excess H$\alpha$
emission EW.

\section{The Ca~{\sc ii} H $\&$ K and H$\epsilon$ lines}

\begin{figure}
\plotone{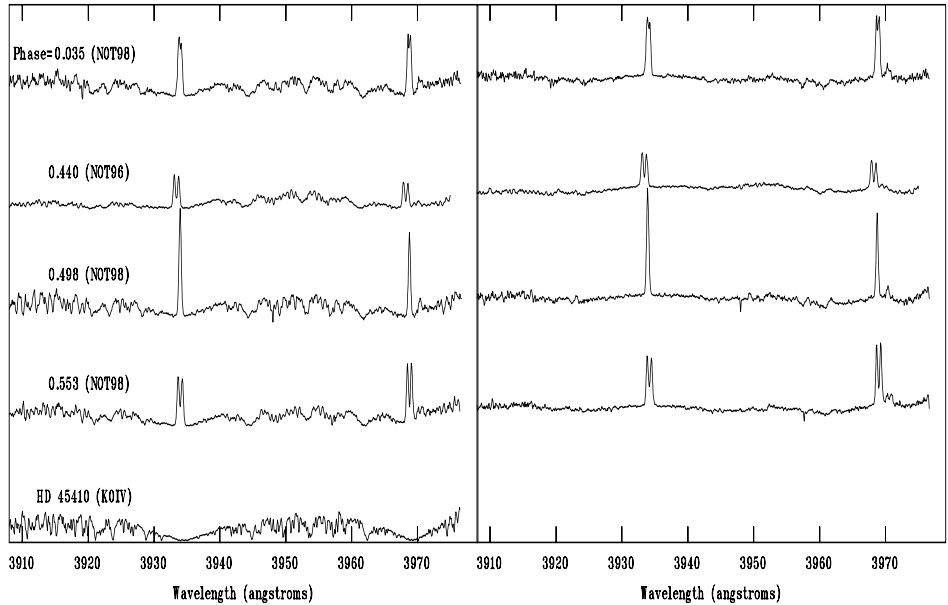}
\caption{As in Fig. 1 for the Ca~{\sc ii} H \& K line region.}
 \label{fig-5}
\end{figure}

We have taken four spectra in the Ca~{\sc ii} H \& K lines region during the
NOT (96 \& 98) observing runs and other spectrum of BF Lyn was taken in
1993 with the 2.2~m telescope at the German Spanish Astronomical
Observatory (CAHA) (Montes et al., 1995). These spectra (Fig.~\ref{fig-5})
 exhibit
clear and strong Ca~{\sc ii} H \& K and H$\epsilon$  emission lines from both
components, in the case of CAHA 93 run the H$\epsilon$ emission line from 
the hot component is overlapped with the Ca~{\sc ii} H emission of the 
cool component.
The excess Ca~{\sc ii} H $\&$ K and H$\epsilon$ emissions change
with the orbital phase during the NOT 98 run in the same way as the
corresponding excess Ca~{\sc ii}  $\lambda$8542 and H$\alpha$ emissions. 
The excess Ca~{\sc ii} H $\&$ K
 EW emissions also show seasonal variations, for instance, the values
of  CAHA 93 observing run are lower than NOT 96 \& 98.

\acknowledgments

This work has been supported by the Universidad Complutense de Madrid
and the Spanish Direcci\'{o}n General de  Ense\~{n}anza Superior e
Investigaci\'{o}n Cient\'{\i}fica (DGESIC) under grant PB97-0259.


\begin{references}

\reference Barden S.C., Nations H.L., 1985, in Cool Stars, Stellar Systems,
and the Sun, ed. M. Zeilik and D.M. Gibson (Springer) p. 262.
\reference Hall J.C., Ramsey L.W., 1992, AJ, 104, 1942
\reference Montes D., De Castro E., Fern\'andez-Figueroa M.J., and Cornide M., 
1995, A\&AS, 114, 287
\reference Strassmeier K.G., Hooten J.T., Hall D.S., and Fekel F.C., 
1989, PASP, 101, 107
\reference Strassmeier  K.G., Hall D.S., Fekel F.C., Scheck M., 
1993, A\&AS, 100, 173
\end{references}
\end{document}